# Performance of the Cherenkov Telescope Array in the presence of clouds

**Mario Pecimotika**[a,b,*], **Katarzyna Adamczyk**[c], **Dijana Dominis Prester**[a], **Orel Gueta**[d], **Dario Hrupec**[e], **Gernot Maier**[c], **Saša Mićanović**[a], **Lovro Pavletić**[a], **Julian Sitarek**[c], **Dorota Sobczyńska**[c] **and Michał Szanecki**[f]

**for the Cherenkov Telescope Array Consortium**
(a complete list of authors can be found at the end of the proceedings)

[a] *Department of Physics, University of Rijeka, Radmile Matejčić 2, Rijeka, Croatia*
[b] *Virovitica University of Applied Sciences, Matije Gupca 78, Virovitica, Croatia*
[c] *Department of Astrophysics, University of Łódź, Pomorska 149, Łódź, Poland*
[d] *Deutsches Elektronen-Synchrotron (DESY), Platanenallee 6, Zeuthen, Germany*
[e] *Department of Physics, J. J. Strossmayer University of Osijek, Trg Ljudevita Gaja 6, Osijek, Croatia*
[f] *Department of Informatics, University of Łódź, Pomorska 149, Łódź, Poland*

*E-mail:* mario.pecimotika@vuv.hr

The Cherenkov Telescope Array (CTA) is the future ground-based observatory for gamma-ray astronomy at very high energies. The atmosphere is an integral part of every Cherenkov telescope. Different atmospheric conditions, such as clouds, can reduce the fraction of Cherenkov photons produced in air showers that reach ground-based telescopes, which may affect the performance. Decreased sensitivity of the telescopes may lead to misconstructed energies and spectra. This study presents the impact of various atmospheric conditions on CTA performance. The atmospheric transmission in a cloudy atmosphere in the wavelength range from 203 nm to 1000 nm was simulated for different cloud bases and different optical depths using the MODerate resolution atmospheric TRANsmission (MODTRAN) code. MODTRAN output files were used as inputs for generic Monte Carlo simulations. The analysis was performed using the MAGIC Analysis and Reconstruction Software (MARS) adapted for CTA. As expected, the effects of clouds are most evident at low energies, near the energy threshold. Even in the presence of dense clouds, high-energy gamma rays may still trigger the telescopes if the first interaction occurs lower in the atmosphere, below the cloud base. A method to analyze very high-energy data obtained in the presence of clouds is presented. The systematic uncertainties of the method are evaluated. These studies help to gain more precise knowledge about the CTA response to cloudy conditions and give insights on how to proceed with data obtained in such conditions. This may prove crucial for alert-based observations and time-critical studies of transient phenomena.

*Presenter

## 1. Introduction

Gamma rays entering the atmosphere interact with the atmospheric nuclei producing cascades of secondary particles, the so-called extensive air showers (EAS). The ultra-relativistic secondary particles in the air shower produce Cherenkov radiation, short flashes of light with the peak at ~ 300 nm in wavelength, that can be recorded using Imaging Air Cherenkov Telescopes (IACT) [1].

Cherenkov Telescope Array (CTA) is the future ground-based observatory for very-high-energy gamma-ray astronomy that will be sensitive in the energy range from 20 GeV to 300 TeV [2, 3]. CTA will consist of large arrays of IACTs located in the Northern (CTA-N, La Palma, Spain,) and Southern (CTA-S, Atacama Desert, Chile) hemispheres. There will be three different types of telescopes, each type covering different energy range: the Large-Sized Telescopes (LST) will provide the low-energy sensitivity; the Medium-Sized Telescopes (MST) will provide the bulk of the sensitivity in the core energy range; and the Small-Sized Telescopes (SST) will provide the high-energy sensitivity.

The atmosphere is an integral part of every IACT and it affects measured Cherenkov light in several ways. First, the number of Cherenkov photons produced in the EAS depends on the refraction index and air density, while the lateral distribution of Cherenkov light depends on the atmospheric profile [4]. Second, Cherenkov photons are scattered and absorbed along the way from the emission point to the detector. The latter is particularly noticeable in the presence of dense clouds and aerosols, which results in a decrease in the number of events near the energy threshold that would otherwise trigger the telescope. Also, the effects of the clouds are not only evident near the energy threshold, but across the entire energy range, reducing an overall trigger yield and biasing events towards lower energies [5].

The impact of the atmospheric conditions on the IACTs has been studied in [6, 7]. An approach to atmospheric corrections for IACTs has been already studied in MAGIC [8] and H.E.S.S. [5, 9] collaborations. In [8] it is shown that for layers of low and moderate optical depths a straightforward correction to the reconstructed energy is possible. The corrected energy is simply obtained as a reconstructed (estimated) energy, $E_{est}$, scaled with the inverse of the average optical depth. In the presence of clouds event reconstruction using simulations based on cloudless atmospheric models [5, 9] yields a bias in the reconstructed energy, therefore the analysis should be performed using simulations that include proper atmospheric conditions.

## 2. Simulation chain

The development of EAS was simulated using `CORSIKA` code version 7.7 [10] with the hadronic interaction model `QGSJET-II`. Simulations have been performed in such a way that they cover low-energy range (4 LSTs, CTA-N), core energy range (15 MSTs, CTA-N), and high-energy range (5 1M-SSTs, CTA-S) for directions of primaries of 20 degrees in Zenith, 180 degrees in Azimuth for CTA-N, and 20 degrees in Zenith, 0 degrees in Azimuth for CTA-S layout (the most favorable directions due to the influence of the geomagnetic field [11]). To simulate the telescope response, CTA Monte Carlo (MC) code, `sim_telarray` with Prod3b settings has been used [12]. The layouts of the telescope systems used are presented in Figure 1.

Because the number of emitted Cherenkov photons is roughly proportional to the primary energy, in an EAS initiated by low-energy gamma rays a smaller number of Cherenkov photons is emitted, which complicates event reconstruction itself and therefore requires more precise and detailed simulations of atmospheric conditions. Hence, the atmospheric transmission in the presence of 1 km thick altostratus clouds, with their bases at 3 and 9 km a.g.l. was simulated for the Northern site (ground altitude at 2174 m a.s.l) using the MODerate resolution atmospheric TRANsmission (`MODTRAN`) code [13]. Two sets of data have been simulated in total, each set differing by the total atmospheric transmission ($T = 0.6, T = 0.8$), in the wavelength range from 200 nm to 1000 nm. `MODTRAN` output was used as an input for `sim_telarray` instead of the default atmospheric file for cloudless conditions.

The atmospheric conditions simulations for the Southern site (ground altitude at 2500 m a.s.l) have also been included within the `sim_telarray`. However, at higher primary energies a significantly larger number of Cherenkov photons is emitted, therefore very precise simulations with `MODTRAN` are not mandatory and the whole process of simulating atmospheric conditions can be simplified. The additional extinction due to cloud presence has been included in the default cloudless atmospheric file for Armazones site using equation (3.11) presented in [14]. Different concentrations of water in cloudy media were used to obtain the total transmissions of $T = 0.8, T = 0.6, T = 0.4$, and $T = 0.2$ in the wavelength range from 250 nm to 700 nm. The clouds at 2.5 and 4.5 km a.g.l. with a thickness of 0.5 km were simulated.

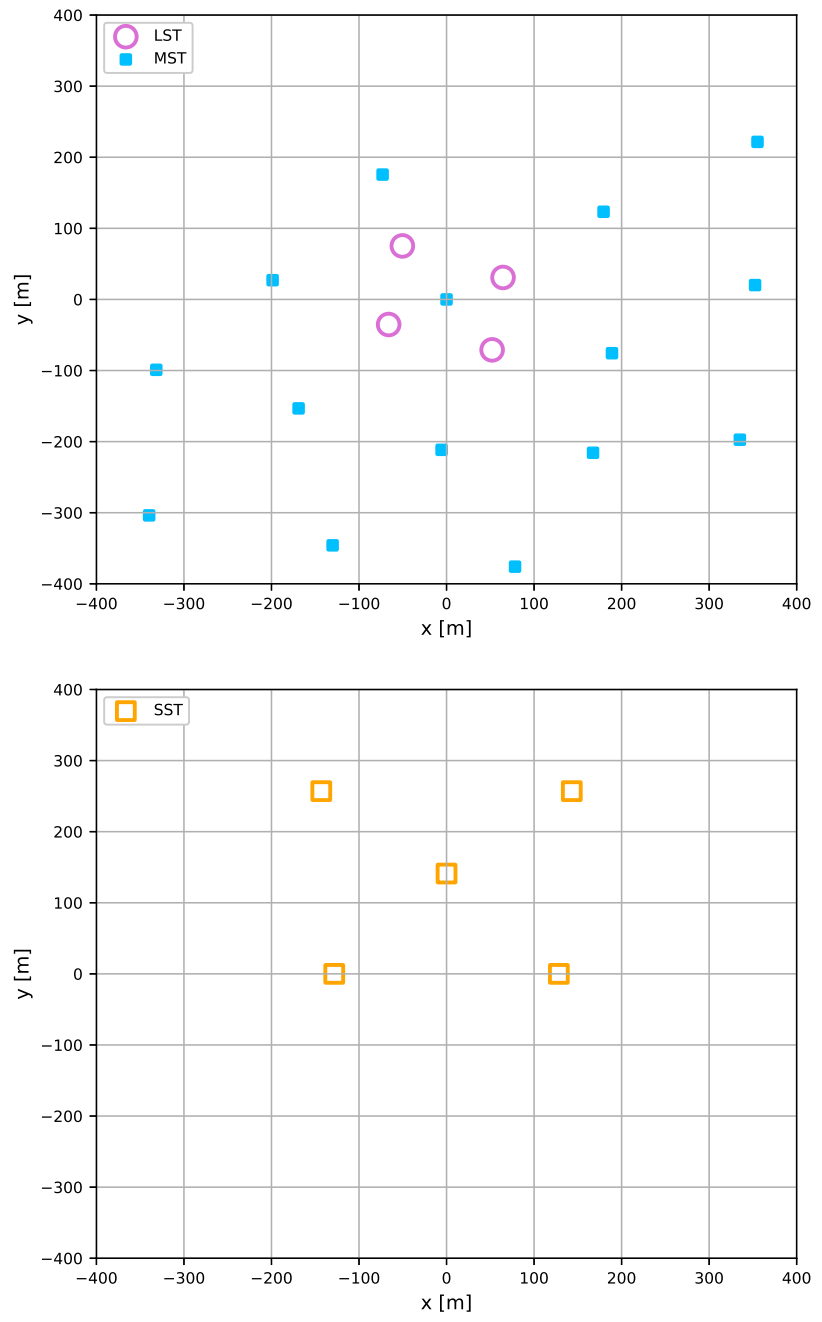

**Figure 1:** Telescope layouts for the Northern (top panel) and Southern (bottom panel) site of the observatory. Note that the layout of the two sites is the one implemented in Prod3.

## 3. Analysis chain

The analysis of the simulated data was performed with MAGIC Analysis and Reconstruction Software (`MARS`) adapted for the CTA use [15, 16]. The full Monte Carlo sample consists of gamma, proton, and electron primary `sim_telarray` subsamples. Before the higher-level analysis, `sim_telarray` files are converted to `ROOT` format using the Convert Hessio Into MARS inPut (`chimp`) package, which performs signal extraction by a *two-pass sliding window* algorithm, image cleaning using the *absolute image cleaning* method and image parametrization. Next, about 5% of the gamma subsample is used to obtain a direction look-up table and to train the telescope-wise energy reconstruction using Random Forest (RF) algorithm [17]. The event direction is obtained as the point in the camera which minimizes the sum of squares of distances to the main axis of the image. Then, another 5% of gamma subsample and 5% of the total proton subsample are used to train gamma/hadron separation using RF, once those subsamples have been processed to the level of stereoscopic reconstruction and energy reconstruction. The rest of the gamma and proton

MC data set, and the full electron subsample are used as a test sample, processed through trained energy and gamma/hadron separation RFs. The final reconstructed energy for each event, $E_{est}$ is calculated as a weighted average of the reconstructed energies of all telescopes, where 1/RMS$^2$ is used as a weight. The global *hadronness*[1] is calculated as a weighted average of the *hadronnesses* calculated by each telescope, where $size_i^{0.54}$ (the $size_i$ is a total number of Cherenkov photons in the camera of the $i$-th telescope) is used as a weight[2]. Finally, gamma-ray selection cuts, $\theta^2$ cut, where $\theta$ is the parameter that specifies the angular distance between the reconstructed and the real source direction, and *hadronness* cut, are optimized and applied to obtain the instrument response functions in means of differential sensitivity, angular resolution, and energy resolution.

## 4. Results and discussion

### 4.1 Differential sensitivity

The differential sensitivity is the minimum flux that can be detected with a statistical significance of $5\sigma$, in this study calculated for 50 hours of observations. The differential sensitivity is calculated in non-overlapping logarithmic energy bins (five per decade), with a minimal number of 10 gamma-ray counts per energy bin, and of signal counts above 5% of the residual background counts.

In Figure 2, the influence of the clouds with $T = 0.6$ and $T = 0.8$ and cloud bases at 3 and 9 km a.g.l. on the differential sensitivity compared to the clear atmosphere ($T = 1.0$) is presented.

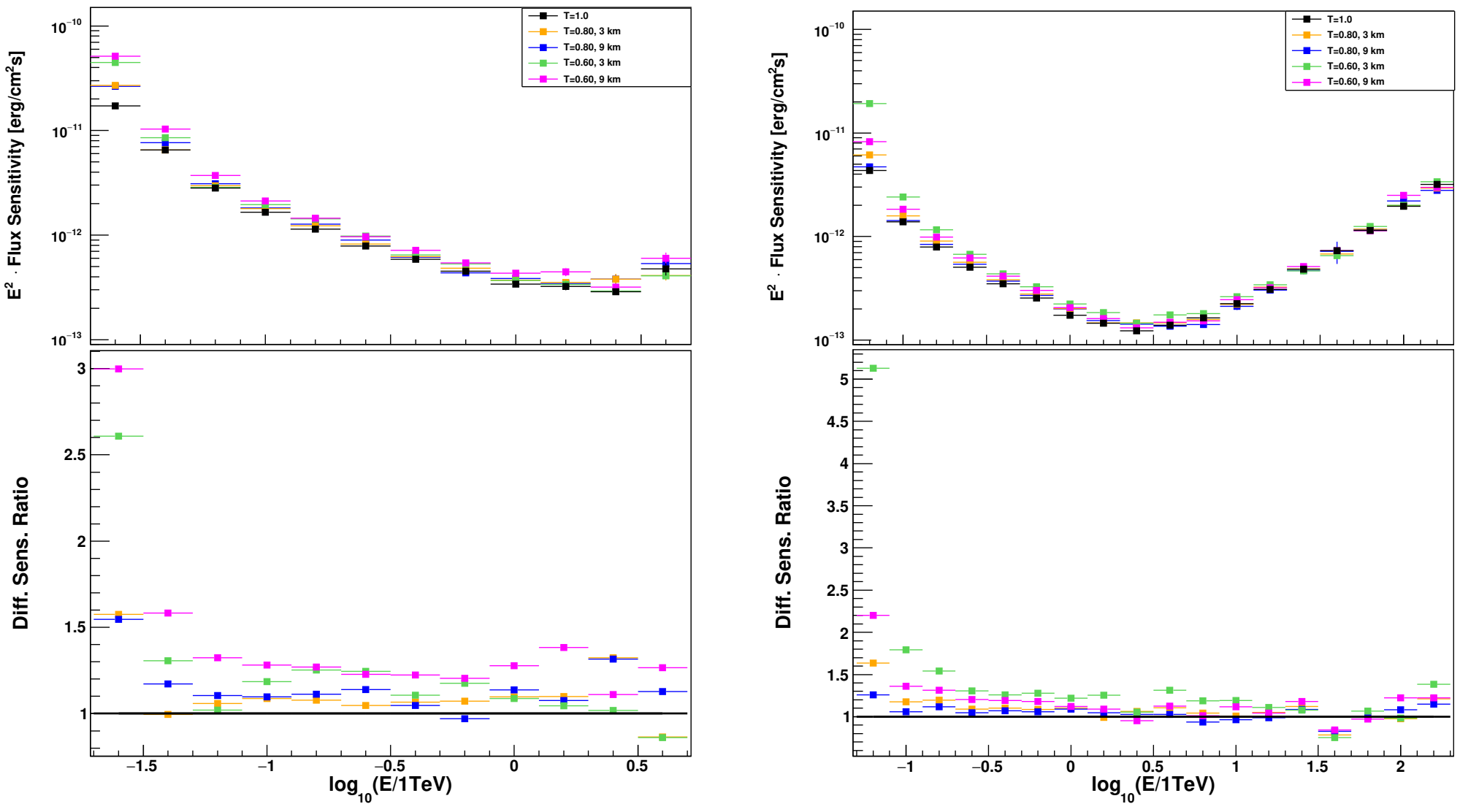

**Figure 2:** The differential sensitivity (**upper panel**) and ratio of differential sensitivities (**bottom panel**) versus $E_{est}$ for the layout of 4 LSTs (**left side**) and 15 MSTs (**right side**).

In the case of 4 LSTs, the most significant impact on the performance of the telescopes have the clouds with the higher cloud base at 9 km a.g.l. and $T = 0.6$, reducing the sensitivity by a factor of ∼ 3 at the energy threshold, with the average reduction of 1.42 in the LSTs sensitivity range. In the case of 15 MSTs, the most prominent impact on the sensitivity of the telescopes have clouds with the lower cloud base at 3 km a.g.l. and the same $T = 0.6$, reducing the sensitivity in the lowest energy bin by a factor of ∼ 5, with the average reduction of 1.44 in the MSTs sensitivity range. An explanation of unexpected behavior that higher clouds of the same $T$ have a greater impact on the sensitivity of the LSTs than lower clouds (which is not observed for MSTs) might be due to the fact that low-energy showers (sensitivity range of LSTs) reach the maximum of development higher in the atmosphere (closer to higher clouds). In the common energy range from 0.1 TeV to 4 TeV for the cloud at 3 km a.g.l. with $T = 0.6$, the average reductions for 4 LSTs and 15 MSTs are, respectively, 1.11 and 1.33. To draw more accurate conclusions further studies are required and planned.

[1]The *hadronness* is a value ranging between 0 and 1 describing how likely the event is initiated by the gamma primary. The closer the *hadronness* is to 0, the event is more gamma-like.

[2]The value of the exponent 0.54 is obtained empirically.

### 4.2 Energy resolution

The assessment of energy reconstruction is obtained by calculating the energy resolution using MC simulations. The energy resolution describes how accurately the instrument can determine the real energy of gamma primary, $E_{true}$, and it is calculated bin-wise as a half width of the interval which contains 68% of the distribution in a respective bin, symmetric around $E_{est}/E_{true} = 1$.

For the layouts of 4 LSTs and 15 MSTs (Figure 3, left and middle panel, respectively), the worst-case scenario appears in the lowest energy bin in the case of clouds with $T = 0.6$, with a reduction in the energy resolution by a factor of ∼ 1.20 for 4 LSTs, and ∼ 1.03 for 15 MSTs.

For large biases in the presence of clouds, the standard definition of the energy resolution is not a useful metric, therefore in the case of 5 SSTs the corrected energy, $E_{cor}$ is used instead of $E_{est}$ (see Section 4.4 or [7] for details). For the layout of 5 SSTs (Figure 3, right panel), the energy resolution reaches its plateau at a value of ∼ 13% in the case of $T \geq 0.6$ and $E_{cor} > 2$ TeV. For lower $E_{cor}$, the energy resolution is poor due to threshold effects.

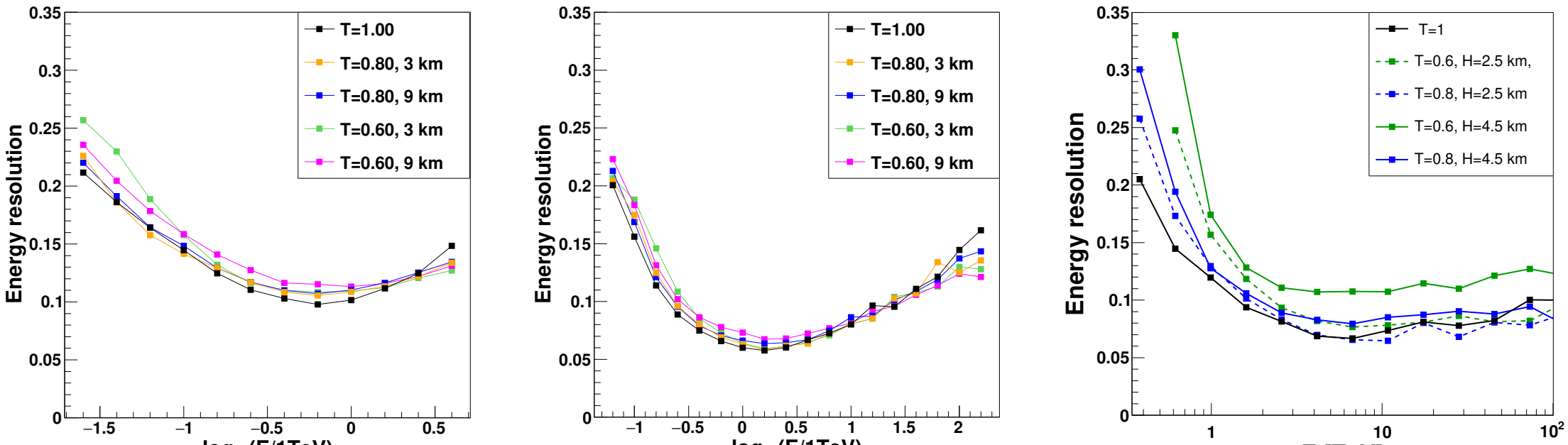

**Figure 3: Left:** The energy resolution versus $E_{est}$ for the layout of 4 LSTs. **Middle:** The energy resolution versus $E_{est}$ for the layout of 15 MSTs. **Right:** The energy resolution versus $E_{cor}$ for the layout of 5 SSTs (taken from [7]).

### 4.3 Angular resolution

The assessment of direction reconstruction is obtained by calculating the angular resolution. The angular resolution is determined from the $\theta$ distribution, and it is represented by the value which contains 68% of all reconstructed gamma-like events in a given energy bin.

In the case of 4 LSTs (Figure 4, left panel), the impact of clouds on the angular resolution is negligible, with largest differences up to ~ 5%. For the layout of 15 MSTs (Figure 4, middle panel), the angular resolution reaches its plateau for $E_{est} > 2.5$ TeV. In that region even in the worst simulated case ($T = 0.6$, 3 km a.g.l.) it amounts to only 0.05 degrees, while the largest difference is ~ 15%. For the layout of 5 SSTs (Figure 4, right panel), the angular resolution is obtained from the MC simulations of the cloudless conditions in such a way that $E_{est}$ is folded with the corrected total atmospheric transmission $\tau_A$ (see Section 4.4 or [7] for details):

$$\sigma_\theta(E_{cor}, T, H) = \sigma_\theta(E_{est} \cdot \tau_A(E_{true}, T, H), 1, 0), \tag{1}$$

where $H$ represent the altitude of the cloud base. In the presence of clouds with $T \geq 0.4$, the angular resolution decreases at $E_{cor} < 4$ TeV, whereupon a plateau is reached. In the case of clouds with $T = 0.2$, the plateau is reached only for $E_{cor} > 10$ TeV.

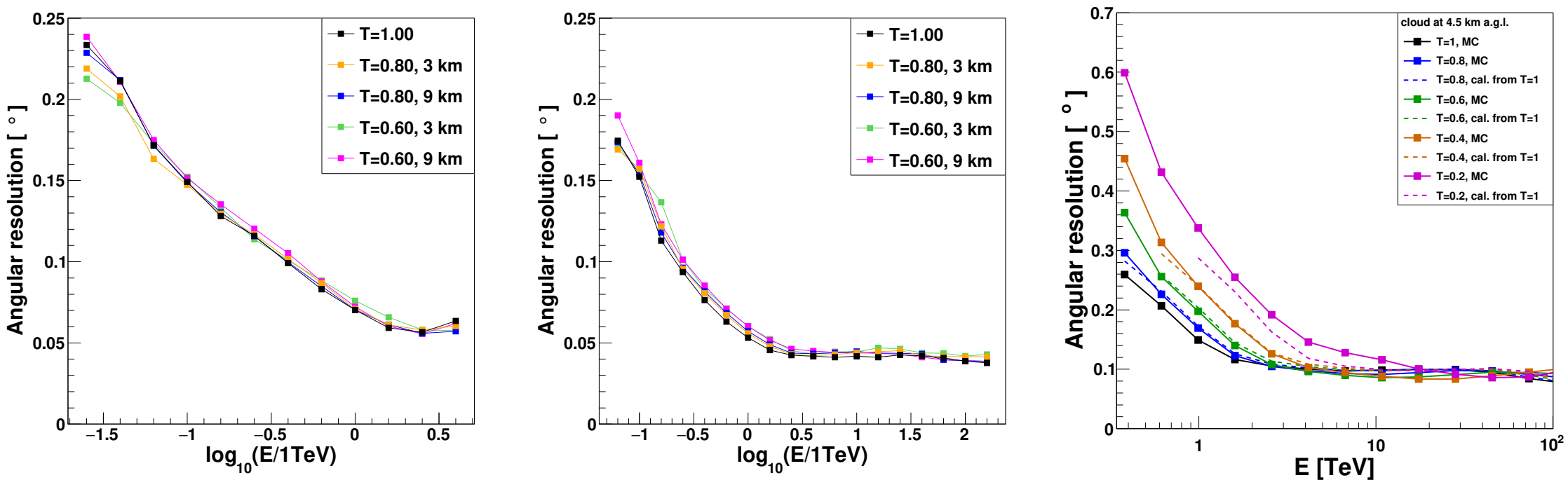

**Figure 4: Left:** The angular resolution versus $E_{est}$ for the layout of 4 LSTs. **Middle:** The angular resolution versus $E_{est}$ for the layout of 15 MSTs. **Right:** The angular resolution versus $E_{cor}$ for the layout of 5 SSTs (note different vertical scale compared to 4 LSTs and 15 MSTs). Dashed lines represent approximation equation (1), while points present the results of the MC simulations. Figure is taken from [7].

### 4.4 Data analysis method for very high energies

To avoid time-consuming MC simulations, a method to analyze high-energy data taken in the presence of clouds ($T < 1.0$) is proposed [7]. The method requires simulations for the cloudless conditions ($T = 0$) and the total atmospheric transmission:

$$\tau(E_{true}, T, H) = 1 - (1 - T) \cdot F_{ab}(E_{true}, H), \tag{2}$$

where $F_{ab}(E_{true}, H)$ is the fraction of Cherenkov photons created above the cloud compared to all photons produced in the EAS, obtained from `CORSIKA` simulations for given energy. To improve the agreement between MC data and this method, a phenomenological correction parameter $A = 1.2$ is introduced [7]:

$$\tau_A(E_{true}, T, H) = 1 - A \cdot \tau(E_{true}, T, H). \tag{3}$$

Relative difference between $E_{est}$ and $E_{true}$ is generally called energy bias. In the presence of clouds, high-energy shower images resemble low-energy events, meaning that $E_{est}$ obtained with standard reconstruction should be corrected [8]. The correction method is based on the energy bias calculated for clear atmosphere, to take into account the energy threshold effects, and $\tau_A(E_{true}, T, H)$. In Figure 5, the results of the method (dashed lines) and energy biases calculated from MC simulations (solid points) are presented. For clouds with cloud bases at 2.5 km and 4.5 km a.g.l., and $T > 0.4$, the bias approximation as proposed in the method [7] may be used to get the corrected energy of reconstructed events. $\tau_A(E_{true}, T, H)$ can be used to reproduce the energy and angular resolutions from MC simulations of cloudless conditions (as shown in previous sections), but also for the assessment of the effective collection area [7].

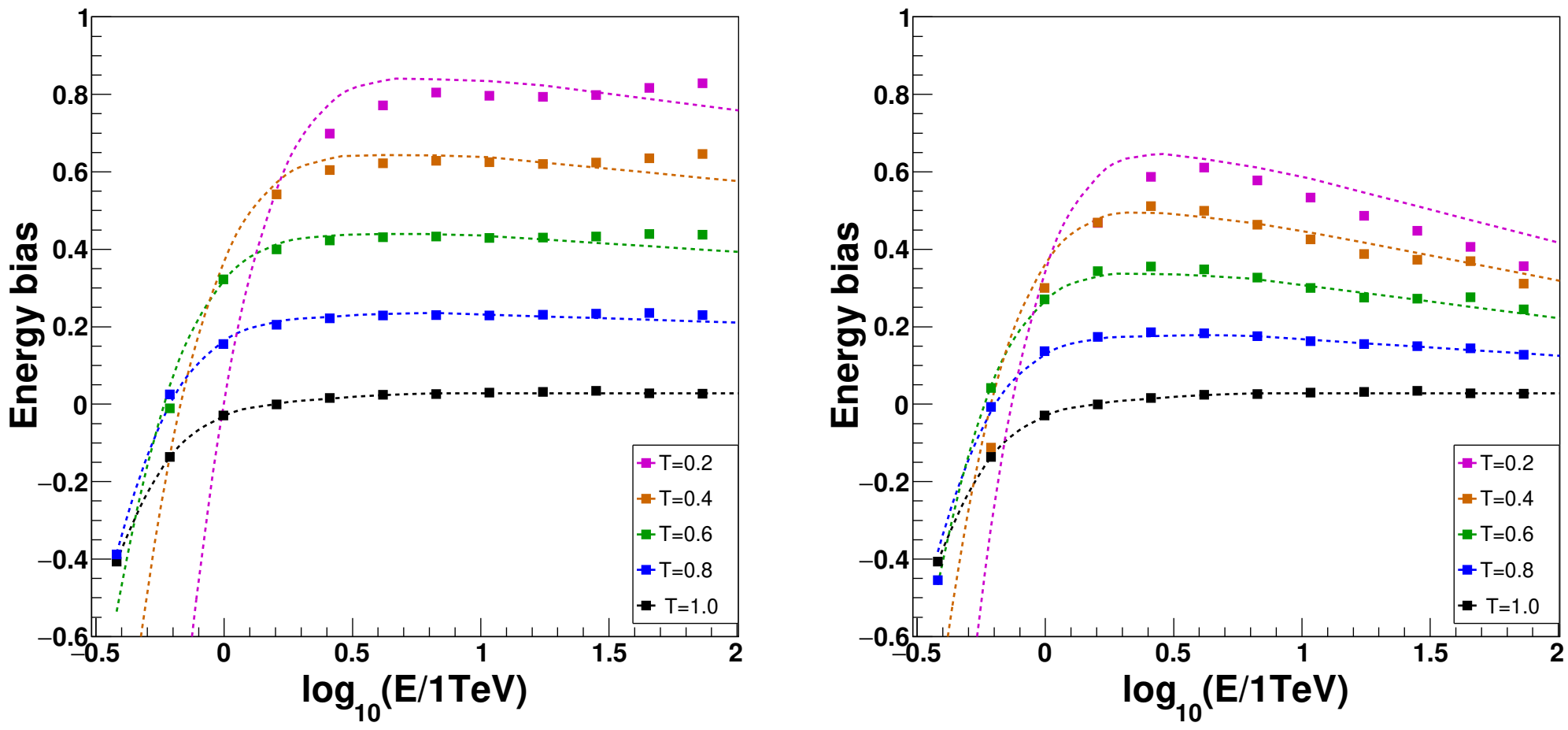

**Figure 5:** The energy bias versus $E_{cor}$ for clouds with cloud bases at 2.5 km a.g.l. (**left panel**) and 4.5 km a.g.l. (**right panel**). Figures are taken from [7].

The spectrum of the potential source can be estimated from the events classified as gamma-like. In the presence of clouds, gamma/hadron separation based on the *hadronness* and $\theta^2$ cuts optimized for clear sky simulations was performed. The reconstructed energy of gamma-like classified events is corrected using the aforementioned method. Then, the flux is calculated in the standard way, but using corrected effective collection area and $E_{cor}$ [7, 8]. In Figure 6, the ratio between the flux reconstructed in the proposed way and the flux obtained from MC simulation for the cloudless conditions (solid and dashed lines, respectively) is presented. In the energy range between 2 TeV and 30 TeV, the expected spectra are underestimated by less than 20% for $T \geq 0.6$. For $E_{cor} > 30$ TeV, the flux underestimation is increased mainly due to the fast degradation of the hadronness cut efficiency. When no corrections are applied (long dashed lines), the underestimation of the spectra is much higher.

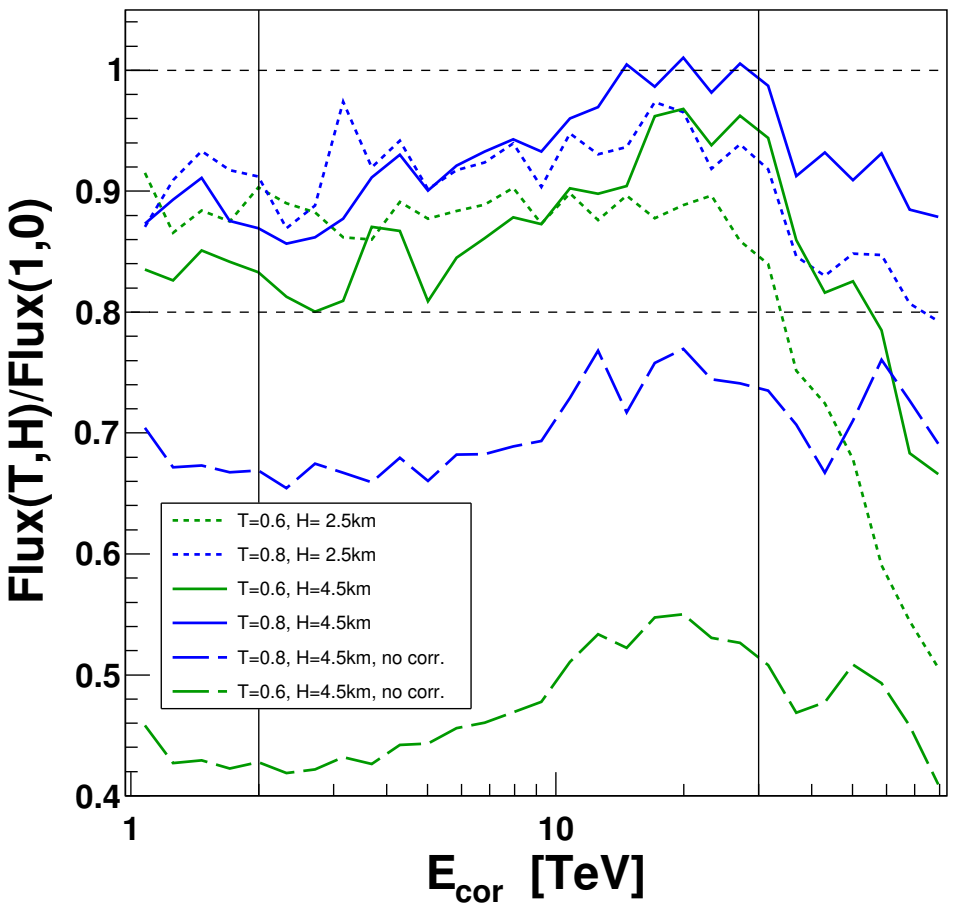

**Figure 6:** The ratio between the reconstructed fluxes. Taken from [7].

## 5. Conclusion

In the presence of high-density clouds with low transmissions through the atmosphere, it is shown that the performance of all three types of CTA telescopes (LSTs, MSTs and SSTs) degrades. Although the degradation effects are most prominent at the energy thresholds, the effects of the clouds are evident across the entire energy range for each telescope type and if not taken into account, they might result in additional systematic errors affecting the measurement.

The effect of the presence of clouds is primarily observed through the reduced number of emitted Cherenkov photons in the shower development, which for low and middle energies (LSTs and MSTs, respectively) leads to the necessity of detailed MC simulations to properly assess the telescope response in the given atmospheric conditions. This may prove crucial for alert-based observations and time-critical studies of transient phenomena at the CTA-N site.

On the other hand, for the high energy range (SSTs foreseen to be built at the CTA-S site) extremely time-consuming MC simulations can be avoided by using a correction method. In this way, the gamma-ray source spectra in the presence of various clouds in the energy range between 2 and 30 TeV have been reconstructed with a systematic uncertainty smaller than $\sim 20\%$ for $T \geq 0.6$. For the lower and upper energy edge of the SSTs sensitivity range, the uncertainty of the method is worse due to threshold effects and the *hadronness* cut efficiency, respectively.

## Acknowledgements

The work performed at the University of Łódź was supported by the Narodowe Centrum Nauki grant No. UMO-2016/22/M/ST9/00583. The work at the University of Rijeka was performed using the supercomputer "Bura" within the Center for Advanced Computing and Modelling. The authors would like to thank CTA Consortium and MAGIC Collaboration for the permission to use their software.

## Full Authors List: Cherenkov Telescope Array Consortium

H. Abdalla[1], H. Abe[2], S. Abe[2], A. Abusleme[3], F. Acero[4], A. Acharyya[5], V. Acín Portella[6], K. Ackley[7], R. Adam[8], C. Adams[9], S.S. Adhikari[10], I. Aguado-Ruesga[11], I. Agudo[12], R. Aguilera[13], A. Aguirre-Santaella[14], F. Aharonian[15], A. Alberdi[12], R. Alfaro[16], J. Alfaro[3], C. Alispach[17], R. Aloisio[18], R. Alves Batista[19], J.-P. Amans[20], L. Amati[21], E. Amato[22], L. Ambrogi[18], G. Ambrosi[23], M. Ambrosio[24], R. Ammendola[25], J. Anderson[26], M. Anduze[8], E.O. Angüner[27], L.A. Antonelli[28], V. Antonuccio[29], P. Antoranz[30], R. Anutarawiramkul[31], J. Aragunde Gutierrez[32], C. Aramo[24], A. Araudo[33,34], M. Araya[35], A. Arbet-Engels[36], C. Arcaro[1], V. Arendt[37], C. Armand[38], T. Armstrong[27], F. Arqueros[11], L. Arrabito[39], B. Arsioli[40], M. Artero[41], K. Asano[2], Y. Ascasíbar[14], J. Aschersleben[42], M. Ashley[43], P. Attinà[44], P. Aubert[45], C. B. Singh[19], D. Baack[46], A. Babic[47], M. Backes[48], V. Baena[13], S. Bajtlik[49], A. Baktash[50], C. Balazs[7], M. Balbo[38], O. Ballester[41], J. Ballet[4], B. Balmaverde[44], A. Bamba[51], R. Bandiera[22], A. Baquero Larriva[11], P. Barai[19], C. Barbier[45], V. Barbosa Martins[52], M. Barcelo[53], M. Barkov[54], M. Barnard[1], L. Baroncelli[21], U. Barres de Almeida[40], J.A. Barrio[11], D. Bastieri[55], P.I. Batista[52], I. Batkovic[55], C. Bauer[53], R. Bautista-González[56], J. Baxter[2], U. Becciani[29], J. Becerra González[32], Y. Becherini[57], G. Beck[58], J. Becker Tjus[59], W. Bednarek[60], A. Belfiore[61], L. Bellizzi[62], R. Belmont[4], W. Benbow[63], D. Berge[52], E. Bernardini[52], M.I. Bernardos[55], K. Bernlöhr[53], A. Berti[64], M. Berton[65], B. Bertucci[23], V. Beshley[66], N. Bhatt[67], S. Bhattacharyya[67], W. Bhattacharyya[52], S. Bhattacharyya[68], B. Bi[69], G. Bicknell[70], N. Biederbeck[46], C. Bigongiari[28], A. Biland[36], R. Bird[71], E. Bissaldi[72], J. Biteau[73], M. Bitossi[74], O. Blanch[41], M. Blank[50], J. Blazek[33], J. Bobin[75], C. Boccato[76], F. Bocchino[77], C. Boehm[78], M. Bohacova[33], C. Boisson[20], J. Boix[41], J.-P. Bolle[52], J. Bolmont[79], G. Bonanno[29], C. Bonavolontà[24], L. Bonneau Arbeletche[80], G. Bonnoli[12], P. Bordas[81], J. Borkowski[49], S. Bórquez[35], R. Bose[82], D. Bose[83], Z. Bosnjak[47], E. Bottacini[55], M. Böttcher[1], M.T. Botticella[84], C. Boutonnet[85], F. Bouyjou[75], V. Bozhilov[86], E. Bozzo[38], L. Brahimi[39], C. Braiding[43], S. Brau-Nogué[87], S. Breen[78], J. Bregeon[39], M. Breuhaus[53], A. Brill[9], W. Brisken[88], E. Brocato[28], A.M. Brown[5], K. Brügge[46], P. Brun[89], P. Brun[39], F. Brun[89], L. Brunetti[45], G. Brunetti[90], P. Bruno[29], A. Bruno[91], A. Bruzzese[6], N. Bucciantini[22], J. Buckley[82], R. Bühler[52], A. Bulgarelli[21], T. Bulik[92], M. Bünning[52], M. Bunse[46], M. Burton[93], A. Burtovoi[76], M. Buscemi[94], S. Buschjäger[46], G. Busetto[55], J. Buss[46], K. Byrum[26], A. Caccianiga[95], F. Cadoux[17], A. Calanducci[29], C. Calderón[3], J. Calvo Tovar[32], R. Cameron[96], P. Campaña[35], R. Canestrari[91], F. Cangemi[79], B. Cantlay[31], M. Capalbi[91], M. Capasso[9], M. Cappi[21], A. Caproni[97], R. Capuzzo-Dolcetta[28], P. Caraveo[61], V. Cárdenas[98], L. Cardiel[41], M. Cardillo[99], C. Carlile[100], S. Caroff[45], R. Carosi[74], A. Carosi[17], E. Carquín[35], M. Carrère[39], J.-M. Casandjian[4], S. Casanova[101,53], E. Cascone[84], F. Cassol[27], A.J. Castro-Tirado[12], F. Catalani[102], O. Catalano[91], D. Cauz[103], A. Ceccanti[64], C. Celestino Silva[80], S. Celli[18], K. Cerny[104], M. Cerruti[85], E. Chabanne[45], P. Chadwick[5], Y. Chai[105], P. Chambery[106], C. Champion[85], S. Chandra[1], S. Chaty[4], A. Chen[58], K. Cheng[2], M. Chernyakova[107], G. Chiaro[61], A. Chiavassa[64,108], M. Chikawa[2], V.R. Chitnis[109], J. Chudoba[33], L. Chytka[104], S. Cikota[47], A. Circiello[24,110], P. Clark[5], M. Çolak[41], E. Colombo[32], J. Colome[13], S. Colonges[85], A. Comastri[21], A. Compagnino[91], V. Conforti[21], E. Congiu[95], R. Coniglione[94], J. Conrad[111], F. Conte[53], J.L. Contreras[11], P. Coppi[112], R. Cornat[8], J. Coronado-Blazquez[14], J. Cortina[113], A. Costa[29], H. Costantini[27], G. Cotter[114], B. Courty[85], S. Covino[95], S. Crestan[61], P. Cristofari[20], R. Crocker[70], J. Croston[115], K. Cubuk[93], O. Cuevas[98], X. Cui[2], G. Cusumano[91], S. Cutini[23], A. D'Aì[91], G. D'Amico[116], F. D'Ammando[90], P. D'Avanzo[95], P. Da Vela[74], M. Dadina[21], S. Dai[117], M. Dalchenko[17], M. Dall' Ora[84], M.K. Daniel[63], J. Dauguet[85], I. Davids[48], J. Davies[114], B. Dawson[118], A. De Angelis[55], A.E. de Araújo Carvalho[40], M. de Bony de Lavergne[45], V. De Caprio[84], G. De Cesare[21], F. De Frondat[20], E.M. de Gouveia Dal Pino[19], I. de la Calle[11], B. De Lotto[103], A. De Luca[61], D. De Martino[84], R.M. de Menezes[19], M. de Naurois[8], E. de Oña Wilhelmi[13], F. De Palma[64], F. De Persio[119], N. de Simone[52], V. de Souza[80], M. Del Santo[91], M.V. del Valle[19], E. Delagnes[75], G. Deleglise[45], M. Delfino Reznicek[6], C. Delgado[113], A.G. Delgado Giler[80], J. Delgado Mengual[6], R. Della Ceca[95], M. Della Valle[84], D. della Volpe[17], D. Depaoli[64,108], D. Depouez[27], J. Devin[85], T. Di Girolamo[24,110], C. Di Giulio[25], A. Di Piano[21], F. Di Pierro[64], L. Di Venere[120], C. Díaz[113], C. Díaz-Bahamondes[3], C. Dib[35], S. Diebold[69], S. Digel[96], R. Dima[55], A. Djannati-Ataï[85], J. Djuvsland[116], A. Dmytriiev[20], K. Docher[9], A. Domínguez[11], D. Dominis Prester[121], A. Donath[53], A. Donini[41], D. Dorner[122], M. Doro[55], R.d.C. dos Anjos[123], J.-L. Dournaux[20], T. Downes[107], G. Drake[26], H. Drass[3], D. Dravins[100], C. Duangchan[31], A. Duara[124], G. Dubus[125], L. Ducci[69], C. Duffy[124], D. Dumora[106], K. Dundas Morå[111], A. Durkalec[126], V.V. Dwarkadas[127], J. Ebr[33], C. Eckner[45], J. Eder[105], A. Ederoclite[19], E. Edy[8], K. Egberts[128], S. Einecke[118], J. Eisch[129], C. Eleftheriadis[130], D. Elsässer[46], G. Emery[17], D. Emmanoulopoulos[115], J.-P. Ernenwein[27], M. Errando[82], P. Escarate[35], J. Escudero[12], C. Espinoza[3], S. Ettori[21], A. Eungwanichayapant[31], P. Evans[124], C. Evoli[18], M. Fairbairn[131], D. Falceta-Goncalves[132], A. Falcone[133], V. Fallah Ramazani[65], R. Falomo[76], K. Farakos[134], G. Fasola[20], A. Fattorini[46], Y. Favre[17], R. Fedora[135], E. Fedorova[136], S. Fegan[8], K. Feijen[118], Q. Feng[9], G. Ferrand[54], G. Ferrara[94], O. Ferreira[8], M. Fesquet[75], E. Fiandrini[23], A. Fiasson[45], M. Filipovic[117], D. Fink[105], J.P. Finley[137], V. Fioretti[21], D.F.G. Fiorillo[24,110], M. Fiorini[61], S. Flis[52], H. Flores[20], L. Foffano[17], C. Föhr[53], M.V. Fonseca[11], L. Font[138], G. Fontaine[8], O. Fornieri[52], P. Fortin[63], L. Fortson[88], N. Fouque[45], A. Fournier[106], B. Fraga[40], A. Franceschini[76], F.J. Franco[30], A. Franco Ordovas[32], L. Freixas Coromina[113], L. Fresnillo[30], C. Fruck[105], D. Fugazza[95], Y. Fujikawa[139], Y. Fujita[2], S. Fukami[2], Y. Fukazawa[140], Y. Fukui[141], D. Fulla[52], S. Funk[142], A. Furniss[143], O. Gabella[39], S. Gabici[85], D. Gaggero[14], G. Galanti[61], G. Galaz[3], P. Galdemard[144], Y. Gallant[39], D. Galloway[7], S. Gallozzi[28], V. Gammaldi[14], R. Garcia[41], E. Garcia[45], E. García[13], R. Garcia López[32], M. Garczarczyk[52], F. Gargano[120], C. Gargano[91], S. Garozzo[29], D. Gascon[81], T. Gasparetto[145], D. Gasparrini[25], H. Gasparyan[52], M. Gaug[138], N. Geffroy[45], A. Gent[146], S. Germani[76], L. Gesa[13], A. Ghalumyan[147], A. Ghedina[148], G. Ghirlanda[95], F. Gianotti[21], S. Giarrusso[91], M. Giarrusso[94], G. Giavitto[52], B. Giebels[8], N. Giglietto[72], V. Gika[134], F. Gillardo[45], R. Gimenes[19], F. Giordano[149], G. Giovannini[90], E. Giro[76], M. Giroletti[90], A. Giuliani[61], L. Giunti[85], M. Gjaja[9], J.-F. Glicenstein[89], P. Gliwny[60], N. Godinovic[150], H. Göksu[53], P. Goldoni[85], J.L. Gómez[12], G. Gómez-Vargas[3], M.M. González[16], J.M. González[151], K.S. Gothe[109], D. Götz[4], J. Goulart Coelho[123], K. Gourgouliatos[5], T. Grabarczyk[152], R. Graciani[81], P. Grandi[21], G. Grasseau[8], D. Grasso[74], A.J. Green[78], D. Green[105], J. Green[28], T. Greenshaw[153], I. Grenier[4], P. Grespan[55], A. Grillo[29], M.-H. Grondin[106], J. Grube[131], V. Guarino[26], B. Guest[37], O. Gueta[52], M. Gündüz[59], S. Gunji[154], A. Gusdorf[20], G. Gyuk[155], J. Hackfeld[59], D. Hadasch[2], J. Haga[139], L. Hagge[52], A. Hahn[105], J.E. Hajlaoui[85], H. Hakobyan[35], A. Halim[89], P. Hamal[33], W. Hanlon[63], S. Hara[156], Y. Harada[157], M.J. Hardcastle[158], M. Harvey[5],

K. Hashiyama[2], T. Hassan Collado[113], T. Haubold[105], A. Haupt[52], U.A. Hautmann[159], M. Havelka[33], K. Hayashi[141], K. Hayashi[160], M. Hayashida[161], H. He[54], L. Heckmann[105], M. Heller[17], J.C. Helo[35], F. Henault[125], G. Henri[125], G. Hermann[53], R. Hermel[45], S. Hernández Cadena[16], J. Herrera Llorente[32], A. Herrero[32], O. Hervet[143], J. Hinton[53], A. Hiramatsu[157], N. Hiroshima[54], K. Hirotani[2], B. Hnatyk[136], R. Hnatyk[136], J.K. Hoang[11], D. Hoffmann[27], W. Hofmann[53], C. Hoischen[128], J. Holder[162], M. Holler[163], B. Hona[164], D. Horan[8], J. Hörandel[165], D. Horns[50], P. Horvath[104], J. Houles[27], T. Hovatta[65], M. Hrabovsky[104], D. Hrupec[166], Y. Huang[135], J.-M. Huet[20], G. Hughes[159], D. Hui[2], G. Hull[73], T.B. Humensky[9], M. Hütten[105], R. Iaria[77], M. Iarlori[18], J.M. Illa[41], R. Imazawa[140], D. Impiombato[91], T. Inada[2], F. Incardona[29], A. Ingallinera[29], Y. Inome[2], S. Inoue[54], T. Inoue[141], Y. Inoue[167], A. Insolia[120,94], F. Iocco[24,110], K. Ioka[168], M. Ionica[23], M. Iori[119], S. Iovenitti[95], A. Iriarte[16], K. Ishio[105], W. Ishizaki[168], Y. Iwamura[2], C. Jablonski[105], J. Jacquemier[45], M. Jacquemont[45], M. Jamrozy[169], P. Janecek[33], F. Jankowsky[170], A. Jardin-Blicq[31], C. Jarnot[87], P. Jean[87], I. Jiménez Martínez[113], W. Jin[171], L. Jocou[125], N. Jordana[172], M. Josselin[73], L. Jouvin[41], I. Jung-Richardt[142], F.J.P.A. Junqueira[19], C. Juramy-Gilles[79], J. Jurysek[38], P. Kaaret[173], L.H.S. Kadowaki[19], M. Kagaya[2], O. Kalekin[142], R. Kankanyan[53], D. Kantzas[174], V. Karas[34], A. Karastergiou[114], S. Karkar[79], E. Kasai[48], J. Kasperek[175], H. Katagiri[176], J. Kataoka[177], K. Katarzyński[178], S. Katsuda[179], U. Katz[142], N. Kawanaka[180], D. Kazanas[130], D. Kerszberg[41], B. Khélifi[85], M.C. Kherlakian[52], T.P. Kian[181], D.B. Kieda[164], T. Kihm[53], S. Kim[3], S. Kimeswenger[163], S. Kisaka[140], R. Kissmann[163], R. Kleijwegt[135], T. Kleiner[52], G. Kluge[10], W. Kluźniak[49], J. Knapp[52], J. Knödlseder[87], A. Kobakhidze[78], Y. Kobayashi[2], B. Koch[3], J. Kocot[152], K. Kohri[182], K. Kokkotas[69], N. Komin[58], A. Kong[2], K. Kosack[4], G. Kowal[132], F. Krack[52], M. Krause[52], F. Krennrich[129], M. Krumholz[70], H. Kubo[180], V. Kudryavtsev[183], S. Kunwar[53], Y. Kuroda[139], J. Kushida[157], P. Kushwaha[19], A. La Barbera[91], N. La Palombara[61], V. La Parola[91], G. La Rosa[91], R. Lahmann[142], G. Lamanna[45], A. Lamastra[28], M. Landoni[95], D. Landriu[4], R.G. Lang[80], J. Lapington[124], P. Laporte[20], P. Lason[152], J. Lasuik[37], J. Lazendic-Galloway[7], T. Le Flour[45], P. Le Sidaner[20], S. Leach[124], A. Leckngam[31], S.-H. Lee[180], W.H. Lee[16], S. Lee[118], M.A. Leigui de Oliveira[184], A. Lemière[85], M. Lemoine-Goumard[106], J.-P. Lenain[79], F. Leone[94,185], V. Leray[8], G. Leto[29], F. Leuschner[69], C. Levy[79,20], R. Lindemann[52], E. Lindfors[65], L. Linhoff[46], I. Liodakis[65], A. Lipniacka[116], S. Lloyd[5], M. Lobo[113], T. Lohse[186], S. Lombardi[28], F. Longo[145], A. Lopez[32], M. López[11], R. López-Coto[55], S. Loporchio[149], F. Louis[75], M. Louys[20], F. Lucarelli[28], D. Lucchesi[55], H. Ludwig Boudi[39], P.L. Luque-Escamilla[56], E. Lyard[38], M.C. Maccarone[91], T. Maccarone[187], E. Mach[101], A.J. Maciejewski[188], J. Mackey[15], G.M. Madejski[96], P. Maeght[39], C. Maggio[138], G. Maier[52], A. Majczyna[126], P. Majumdar[83,2], M. Makariev[189], M. Mallamaci[55], R. Malta Nunes de Almeida[184], S. Maltezos[134], D. Malyshev[142], D. Malyshev[69], D. Mandat[33], G. Maneva[189], M. Manganaro[121], G. Manicò[94], P. Manigot[8], K. Mannheim[122], N. Maragos[134], D. Marano[29], M. Marconi[84], A. Marcowith[39], M. Marculewicz[190], B. Marčun[68], J. Marín[98], N. Marinello[55], P. Marinos[118], M. Mariotti[55], S. Markoff[174], P. Marquez[41], G. Marsella[94], J. Martí[56], J.-M. Martin[20], P. Martin[87], O. Martinez[30], M. Martínez[41], G. Martínez[113], O. Martínez[41], H. Martínez-Huerta[80], C. Marty[87], R. Marx[53], N. Masetti[21,151], P. Massimino[29], A. Mastichiadis[191], H. Matsumoto[167], N. Matthews[164], G. Maurin[45], W. Max-Moerbeck[192], N. Maxted[43], D. Mazin[2,105], M.N. Mazziotta[120], S.M. Mazzola[77], J.D. Mbarubucyeye[52], L. Mc Comb[5], I. McHardy[115], S. McKeague[107], S. McMuldroch[63], E. Medina[64], D. Medina Miranda[17], A. Melandri[95], C. Melioli[19], D. Melkumyan[52], S. Menchiari[62], S. Mender[46], S. Mereghetti[61], G. Merino Arévalo[6], E. Mestre[13], J.-L. Meunier[79], T. Meures[135], M. Meyer[142], S. Micanovic[121], M. Miceli[77], M. Michailidis[69], J. Michałowski[101], T. Miener[11], I. Mievre[45], J. Miller[35], I.A. Minaya[153], T. Mineo[91], M. Minev[189], J.M. Miranda[30], R. Mirzoyan[105], A. Mitchell[36], T. Mizuno[193], B. Mode[135], R. Moderski[49], L. Mohrmann[142], E. Molina[81], E. Molinari[148], T. Montaruli[17], I. Monteiro[45], C. Moore[124], A. Moralejo[41], D. Morcuende-Parrilla[11], E. Moretti[41], L. Morganti[64], K. Mori[194], P. Moriarty[15], K. Morik[46], G. Morlino[22], P. Morris[114], A. Morselli[25], K. Mosshammer[52], P. Moya[192], R. Mukherjee[9], J. Muller[8], C. Mundell[172], J. Mundet[41], T. Murach[52], A. Muraczewski[49], H. Muraishi[195], K. Murase[2], I. Musella[84], A. Musumarra[120], A. Nagai[17], N. Nagar[196], S. Nagataki[54], T. Naito[156], T. Nakamori[154], K. Nakashima[142], K. Nakayama[51], N. Nakhjiri[13], G. Naletto[55], D. Naumann[52], L. Nava[95], R. Navarro[174], M.A. Nawaz[132], H. Ndiyavala[1], D. Neise[36], L. Nellen[16], R. Nemmen[19], M. Newbold[164], N. Neyroud[45], K. Ngernphat[31], T. Nguyen Trung[73], L. Nicastro[21], L. Nickel[46], J. Niemiec[101], D. Nieto[11], M. Nievas[32], C. Nigro[41], M. Nikołajuk[190], D. Ninci[41], K. Nishijima[157], K. Noda[2], Y. Nogami[176], S. Nolan[5], R. Nomura[2], R. Norris[117], D. Nosek[197], M. Nöthe[46], B. Novosyadlyj[198], V. Novotny[197], S. Nozaki[180], F. Nunio[144], P. O'Brien[124], K. Obara[176], R. Oger[85], Y. Ohira[51], M. Ohishi[2], S. Ohm[52], Y. Ohtani[2], T. Oka[180], N. Okazaki[2], A. Okumura[139,199], J.-F. Olive[87], C. Oliver[30], G. Olivera[52], B. Olmi[22], R.A. Ong[71], M. Orienti[90], R. Orito[200], M. Orlandini[21], S. Orlando[77], E. Orlando[145], J.P. Osborne[124], M. Ostrowski[169], N. Otte[146], E. Ovcharov[86], E. Owen[2], I. Oya[159], A. Ozieblo[152], M. Padovani[22], I. Pagano[29], A. Pagliaro[91], A. Paizis[61], M. Palatiello[145], M. Palatka[33], E. Palazzi[21], J.-L. Panazol[45], D. Paneque[105], B. Panes[3], S. Panny[163], F.R. Pantaleo[72], M. Panter[53], R. Paoletti[62], M. Paolillo[24,110], A. Papitto[28], A. Paravac[122], J.M. Paredes[81], G. Pareschi[95], N. Park[127], N. Parmiggiani[21], R.D. Parsons[186], P. Paśko[201], S. Patel[52], B. Patricelli[28], G. Pauletta[103], L. Pavletić[121], S. Pavy[8], A. Pe'er[105], M. Pech[33], M. Pecimotika[121], M.G. Pellegriti[120], P. Peñil Del Campo[11], M. Penno[52], A. Pepato[55], S. Perard[106], C. Perennes[55], G. Peres[77], M. Peresano[4], A. Pérez-Aguilera[11], J. Pérez-Romero[14], M.A. Pérez-Torres[12], M. Perri[28], M. Persic[103], S. Petrera[18], P.-O. Petrucci[125], O. Petruk[66], B. Peyaud[89], K. Pfrang[52], E. Pian[21], G. Piano[99], P. Piatteli[94], E. Pietropaolo[18], R. Pillera[149], B. Pilszyk[101], D. Pimentel[202], F. Pintore[91], C. Pio García[41], G. Pirola[64], F. Piron[39], A. Pisarski[190], S. Pita[85], M. Pohl[128], V. Poireau[45], P. Poledrelli[159], A. Pollo[126], M. Polo[113], C. Pongkitivanichkul[31], J. Porthault[144], J. Powell[171], D. Pozo[98], R.R. Prado[52], E. Prandini[55], P. Prasit[31], J. Prast[45], K. Pressard[73], G. Principe[90], C. Priyadarshi[41], N. Produit[38], D. Prokhorov[174], H. Prokoph[52], M. Prouza[33], H. Przybilski[101], E. Pueschel[52], G. Pühlhofer[69], I. Puljak[150], M.L. Pumo[94], M. Punch[85,57], F. Queiroz[203], J. Quinn[204], A. Quirrenbach[170], S. Rainò[149], P.J. Rajda[175], R. Rando[55], S. Razzaque[205], E. Rebert[20], S. Recchia[85], P. Reichherzer[59], O. Reimer[163], A. Reimer[163], A. Reisenegger[3,206], Q. Remy[53], M. Renaud[39], T. Reposeur[106], B. Reville[53], J.-M. Reymond[75], J. Reynolds[15], W. Rhode[46], D. Ribeiro[9], M. Ribó[81], G. Richards[162], T. Richtler[196], J. Rico[41], F. Rieger[53], L. Riitano[135], V. Ripepi[84], M. Riquelme[192], D. Riquelme[35], S. Rivoire[39], V. Rizi[18], E. Roache[63], B. Röben[159], M. Roche[106], J. Rodriguez[4], G. Rodriguez Fernandez[25], J.C. Rodriguez Ramirez[19], J.J. Rodríguez Vázquez[113], F. Roepke[170], G. Rojas[207], L. Romanato[55], P. Romano[95], G. Romeo[29], F. Romero Lobato[11], C. Romoli[53], M. Roncadelli[103], S. Ronda[30], J. Rosado[11], A. Rosales de Leon[5], G. Rowell[118], B. Rudak[49], A. Rugliancich[74], J.E. Ruíz del Mazo[12], W. Rujopakarn[31], C. Rulten[5], C. Russell[3],

F. Russo[21], I. Sadeh[52], E. Sæther Hatlen[10], S. Safi-Harb[37], L. Saha[11], P. Saha[208], V. Sahakian[147], S. Sailer[53], T. Saito[2], N. Sakaki[54], S. Sakurai[2], F. Salesa Greus[101], G. Salina[25], H. Salzmann[69], D. Sanchez[45], M. Sánchez-Conde[14], H. Sandaker[10], A. Sandoval[16], P. Sangiorgi[91], M. Sanguillon[39], H. Sano[2], M. Santander[171], A. Santangelo[69], E.M. Santos[202], R. Santos-Lima[19], A. Sanuy[81], L. Sapozhnikov[96], T. Saric[150], S. Sarkar[114], H. Sasaki[157], N. Sasaki[179], K. Satalecka[52], Y. Sato[209], F.G. Saturni[28], M. Sawada[54], U. Sawangwit[31], J. Schaefer[142], A. Scherer[3], J. Scherpenberg[105], P. Schipani[84], B. Schleicher[122], J. Schmoll[5], M. Schneider[143], H. Schoorlemmer[53], P. Schovanek[33], F. Schussler[89], B. Schwab[142], U. Schwanke[186], J. Schwarz[95], T. Schweizer[105], E. Sciacca[29], S. Scuderi[61], M. Seglar Arroyo[45], A. Segreto[91], I. Seitenzahl[43], D. Semikoz[85], O. Sergijenko[136], J.E. Serna Franco[16], M. Servillat[20], K. Seweryn[201], V. Sguera[21], A. Shalchi[37], R.Y. Shang[71], P. Sharma[73], R.C. Shellard[40], L. Sidoli[61], J. Sieiro[81], H. Siejkowski[152], J. Silk[114], A. Sillanpää[65], B.B. Singh[109], K.K. Singh[210], A. Sinha[39], C. Siqueira[80], G. Sironi[95], J. Sitarek[60], P. Sizun[75], V. Sliusar[38], A. Slowikowska[178], D. Sobczyńska[60], R.W. Sobrinho[184], H. Sol[20], G. Sottile[91], H. Spackman[114], A. Specovius[142], S. Spencer[114], G. Spengler[186], D. Spiga[95], A. Spolon[55], W. Springer[164], A. Stamerra[28], S. Stanič[68], R. Starling[124], Ł. Stawarz[169], R. Steenkamp[48], S. Stefanik[197], C. Stegmann[128], A. Steiner[52], S. Steinmassl[53], C. Stella[103], C. Steppa[128], R. Sternberger[52], M. Sterzel[152], C. Stevens[135], B. Stevenson[71], T. Stolarczyk[4], G. Stratta[21], U. Straumann[208], J. Striškovič[166], M. Strzys[2], R. Stuik[174], M. Suchenek[211], Y. Suda[140], Y. Sunada[179], T. Suomijarvi[73], T. Suric[212], P. Sutcliffe[153], H. Suzuki[213], P. Świerk[101], T. Szepieniec[152], A. Tacchini[21], K. Tachihara[141], G. Tagliaferri[95], H. Tajima[139], N. Tajima[2], D. Tak[52], K. Takahashi[214], H. Takahashi[140], M. Takahashi[2], M. Takahashi[2], J. Takata[2], R. Takeishi[2], T. Tam[2], M. Tanaka[182], T. Tanaka[213], S. Tanaka[209], D. Tateishi[179], M. Tavani[99], F. Tavecchio[95], T. Tavernier[89], L. Taylor[135], A. Taylor[52], L.A. Tejedor[11], P. Temnikov[189], Y. Terada[179], K. Terauchi[180], J.C. Terrazas[192], R. Terrier[85], T. Terzic[121], M. Teshima[105,2], V. Testa[28], D. Thibaut[85], F. Thocquenne[75], W. Tian[2], L. Tibaldo[87], A. Tiengo[215], D. Tiziani[142], M. Tluczykont[50], C.J. Todero Peixoto[102], F. Tokanai[154], K. Toma[160], L. Tomankova[142], J. Tomastik[104], D. Tonev[189], M. Tornikoski[216], D.F. Torres[13], E. Torresi[21], G. Tosti[95], L. Tosti[23], T. Totani[51], N. Tothill[117], F. Toussenel[79], G. Tovmassian[16], P. Travnicek[33], C. Trichard[8], M. Trifoglio[21], A. Trois[95], S. Truzzi[62], A. Tsiahina[87], T. Tsuru[180], B. Turk[45], A. Tutone[91], Y. Uchiyama[161], G. Umana[29], P. Utayarat[31], L. Vaclavek[104], M. Vacula[104], V. Vagelli[23,217], F. Vagnetti[25], F. Vakili[218], J.A. Valdivia[192], M. Valentino[24], A. Valio[19], B. Vallage[89], P. Vallania[44,64], J.V. Valverde Quispe[8], A.M. Van den Berg[42], W. van Driel[20], C. van Eldik[142], C. van Rensburg[1], B. van Soelen[210], J. Vandenbroucke[135], J. Vanderwalt[1], G. Vasileiadis[39], V. Vassiliev[71], M. Vázquez Acosta[32], M. Vecchi[42], A. Vega[98], J. Veh[142], P. Veitch[118], P. Venault[75], C. Venter[1], S. Ventura[62], S. Vercellone[95], S. Vergani[20], V. Verguilov[189], G. Verna[27], S. Vernetto[44,64], V. Verzi[25], G.P. Vettolani[90], C. Veyssiere[144], I. Viale[55], A. Viana[80], N. Viaux[35], J. Vicha[33], J. Vignatti[35], C.F. Vigorito[64,108], J. Villanueva[98], J. Vink[174], V. Vitale[23], V. Vittorini[99], V. Vodeb[68], H. Voelk[53], N. Vogel[142], V. Voisin[79], S. Vorobiov[68], I. Vovk[2], M. Vrastil[33], T. Vuillaume[45], S.J. Wagner[170], R. Wagner[105], P. Wagner[52], K. Wakazono[139], S.P. Wakely[127], R. Walter[38], M. Ward[5], D. Warren[54], J. Watson[52], N. Webb[87], M. Wechakama[31], P. Wegner[52], A. Weinstein[129], C. Weniger[174], F. Werner[53], H. Wetteskind[105], M. White[118], R. White[53], A. Wierzcholska[101], S. Wiesand[52], R. Wijers[174], M. Wilkinson[124], M. Will[105], D.A. Williams[143], J. Williams[124], T. Williamson[162], A. Wolter[95], Y.W. Wong[142], M. Wood[96], C. Wunderlich[62], T. Yamamoto[213], H. Yamamoto[141], Y. Yamane[141], R. Yamazaki[209], S. Yanagita[176], L. Yang[205], S. Yoo[180], T. Yoshida[176], T. Yoshikoshi[2], P. Yu[71], P. Yu[85], A. Yusafzai[59], M. Zacharias[20], G. Zaharijas[68], B. Zaldivar[14], L. Zampieri[76], R. Zanmar Sanchez[29], D. Zaric[150], M. Zavrtanik[68], D. Zavrtanik[68], A.A. Zdziarski[49], A. Zech[20], H. Zechlin[64], A. Zenin[139], A. Zerwekh[35], V.I. Zhdanov[136], K. Ziętara[169], A. Zink[142], J. Ziółkowski[49], V. Zitelli[21], M. Živec[68], A. Zmija[142]

1: Centre for Space Research, North-West University, Potchefstroom, 2520, South Africa

2: Institute for Cosmic Ray Research, University of Tokyo, 5-1-5, Kashiwa-no-ha, Kashiwa, Chiba 277-8582, Japan

3: Pontificia Universidad Católica de Chile, Av. Libertador Bernardo O'Higgins 340, Santiago, Chile

4: AIM, CEA, CNRS, Université Paris-Saclay, Université Paris Diderot, Sorbonne Paris Cité, CEA Paris-Saclay, IRFU/DAp, Bat 709, Orme des Merisiers, 91191 Gif-sur-Yvette, France

5: Centre for Advanced Instrumentation, Dept. of Physics, Durham University, South Road, Durham DH1 3LE, United Kingdom

6: Port d'Informació Científica, Edifici D, Carrer de l'Albareda, 08193 Bellaterrra (Cerdanyola del Vallès), Spain

7: School of Physics and Astronomy, Monash University, Melbourne, Victoria 3800, Australia

8: Laboratoire Leprince-Ringuet, École Polytechnique (UMR 7638, CNRS/IN2P3, Institut Polytechnique de Paris), 91128 Palaiseau, France

9: Department of Physics, Columbia University, 538 West 120th Street, New York, NY 10027, USA

10: University of Oslo, Department of Physics, Sem Saelandsvei 24 - PO Box 1048 Blindern, N-0316 Oslo, Norway

11: EMFTEL department and IPARCOS, Universidad Complutense de Madrid, 28040 Madrid, Spain

12: Instituto de Astrofísica de Andalucía-CSIC, Glorieta de la Astronomía s/n, 18008, Granada, Spain

13: Institute of Space Sciences (ICE-CSIC), and Institut d'Estudis Espacials de Catalunya (IEEC), and Institució Catalana de Recerca I Estudis Avançats (ICREA), Campus UAB, Carrer de Can Magrans, s/n 08193 Cerdanyola del Vallés, Spain

14: Instituto de Física Teórica UAM/CSIC and Departamento de Física Teórica, Universidad Autónoma de Madrid, c/ Nicolás Cabrera 13-15, Campus de Cantoblanco UAM, 28049 Madrid, Spain

15: Dublin Institute for Advanced Studies, 31 Fitzwilliam Place, Dublin 2, Ireland

16: Universidad Nacional Autónoma de México, Delegación Coyoacán, 04510 Ciudad de México, Mexico

17: University of Geneva - Département de physique nucléaire et corpusculaire, 24 rue du Général-Dufour, 1211 Genève 4, Switzerland

18: INFN Dipartimento di Scienze Fisiche e Chimiche - Università degli Studi dell'Aquila and Gran Sasso Science Institute, Via Vetoio 1, Viale Crispi 7, 67100 L'Aquila, Italy

19: Instituto de Astronomia, Geofísico, e Ciências Atmosféricas - Universidade de São Paulo, Cidade Universitária, R. do Matão, 1226, CEP 05508-090, São Paulo, SP, Brazil 20: LUTH, GEPI and LERMA, Observatoire de Paris, CNRS, PSL University, 5 place Jules Janssen, 92190, Meudon, France
21: INAF - Osservatorio di Astrofisica e Scienza dello spazio di Bologna, Via Piero Gobetti 93/3, 40129 Bologna, Italy
22: INAF - Osservatorio Astrofisico di Arcetri, Largo E. Fermi, 5 - 50125 Firenze, Italy
23: INFN Sezione di Perugia and Università degli Studi di Perugia, Via A. Pascoli, 06123 Perugia, Italy
24: INFN Sezione di Napoli, Via Cintia, ed. G, 80126 Napoli, Italy
25: INFN Sezione di Roma Tor Vergata, Via della Ricerca Scientifica 1, 00133 Rome, Italy
26: Argonne National Laboratory, 9700 S. Cass Avenue, Argonne, IL 60439, USA
27: Aix-Marseille Université, CNRS/IN2P3, CPPM, 163 Avenue de Luminy, 13288 Marseille cedex 09, France
28: INAF - Osservatorio Astronomico di Roma, Via di Frascati 33, 00040, Monteporzio Catone, Italy
29: INAF - Osservatorio Astrofisico di Catania, Via S. Sofia, 78, 95123 Catania, Italy
30: Grupo de Electronica, Universidad Complutense de Madrid, Av. Complutense s/n, 28040 Madrid, Spain
31: National Astronomical Research Institute of Thailand, 191 Huay Kaew Rd., Suthep, Muang, Chiang Mai, 50200, Thailand
32: Instituto de Astrofísica de Canarias and Departamento de Astrofísica, Universidad de La Laguna, La Laguna, Tenerife, Spain
33: FZU - Institute of Physics of the Czech Academy of Sciences, Na Slovance 1999/2, 182 21 Praha 8, Czech Republic
34: Astronomical Institute of the Czech Academy of Sciences, Bocni II 1401 - 14100 Prague, Czech Republic
35: CCTVal, Universidad Técnica Federico Santa María, Avenida España 1680, Valparaíso, Chile
36: ETH Zurich, Institute for Particle Physics, Schafmattstr. 20, CH-8093 Zurich, Switzerland
37: The University of Manitoba, Dept of Physics and Astronomy, Winnipeg, Manitoba R3T 2N2, Canada
38: Department of Astronomy, University of Geneva, Chemin d'Ecogia 16, CH-1290 Versoix, Switzerland
39: Laboratoire Univers et Particules de Montpellier, Université de Montpellier, CNRS/IN2P3, CC 72, Place Eugène Bataillon, F-34095 Montpellier Cedex 5, France
40: Centro Brasileiro de Pesquisas Físicas, Rua Xavier Sigaud 150, RJ 22290-180, Rio de Janeiro, Brazil
41: Institut de Fisica d'Altes Energies (IFAE), The Barcelona Institute of Science and Technology, Campus UAB, 08193 Bellaterra (Barcelona), Spain
42: University of Groningen, KVI - Center for Advanced Radiation Technology, Zernikelaan 25, 9747 AA Groningen, The Netherlands
43: School of Physics, University of New South Wales, Sydney NSW 2052, Australia
44: INAF - Osservatorio Astrofisico di Torino, Strada Osservatorio 20, 10025 Pino Torinese (TO), Italy
45: Univ. Savoie Mont Blanc, CNRS, Laboratoire d'Annecy de Physique des Particules - IN2P3, 74000 Annecy, France
46: Department of Physics, TU Dortmund University, Otto-Hahn-Str. 4, 44221 Dortmund, Germany
47: University of Zagreb, Faculty of electrical engineering and computing, Unska 3, 10000 Zagreb, Croatia
48: University of Namibia, Department of Physics, 340 Mandume Ndemufayo Ave., Pioneerspark, Windhoek, Namibia
49: Nicolaus Copernicus Astronomical Center, Polish Academy of Sciences, ul. Bartycka 18, 00-716 Warsaw, Poland
50: Universität Hamburg, Institut für Experimentalphysik, Luruper Chaussee 149, 22761 Hamburg, Germany
51: Graduate School of Science, University of Tokyo, 7-3-1 Hongo, Bunkyo-ku, Tokyo 113-0033, Japan
52: Deutsches Elektronen-Synchrotron, Platanenallee 6, 15738 Zeuthen, Germany
53: Max-Planck-Institut für Kernphysik, Saupfercheckweg 1, 69117 Heidelberg, Germany
54: RIKEN, Institute of Physical and Chemical Research, 2-1 Hirosawa, Wako, Saitama, 351-0198, Japan
55: INFN Sezione di Padova and Università degli Studi di Padova, Via Marzolo 8, 35131 Padova, Italy
56: Escuela Politécnica Superior de Jaén, Universidad de Jaén, Campus Las Lagunillas s/n, Edif. A3, 23071 Jaén, Spain
57: Department of Physics and Electrical Engineering, Linnaeus University, 351 95 Växjö, Sweden
58: University of the Witwatersrand, 1 Jan Smuts Avenue, Braamfontein, 2000 Johannesburg, South Africa
59: Institut für Theoretische Physik, Lehrstuhl IV: Plasma-Astroteilchenphysik, Ruhr-Universität Bochum, Universitätsstraße 150, 44801 Bochum, Germany
60: Faculty of Physics and Applied Computer Science, University of Lódź, ul. Pomorska 149-153, 90-236 Lódź, Poland
61: INAF - Istituto di Astrofisica Spaziale e Fisica Cosmica di Milano, Via A. Corti 12, 20133 Milano, Italy
62: INFN and Università degli Studi di Siena, Dipartimento di Scienze Fisiche, della Terra e dell'Ambiente (DSFTA), Sezione di Fisica, Via Roma 56, 53100 Siena, Italy
63: Center for Astrophysics | Harvard & Smithsonian, 60 Garden St, Cambridge, MA 02180, USA
64: INFN Sezione di Torino, Via P. Giuria 1, 10125 Torino, Italy
65: Finnish Centre for Astronomy with ESO, University of Turku, Finland, FI-20014 University of Turku, Finland
66: Pidstryhach Institute for Applied Problems in Mechanics and Mathematics NASU, 3B Naukova Street, Lviv, 79060, Ukraine
67: Bhabha Atomic Research Centre, Trombay, Mumbai 400085, India
68: Center for Astrophysics and Cosmology, University of Nova Gorica, Vipavska 11c, 5270 Ajdovščina, Slovenia
69: Institut für Astronomie und Astrophysik, Universität Tübingen, Sand 1, 72076 Tübingen, Germany
70: Research School of Astronomy and Astrophysics, Australian National University, Canberra ACT 0200, Australia
71: Department of Physics and Astronomy, University of California, Los Angeles, CA 90095, USA
72: INFN Sezione di Bari and Politecnico di Bari, via Orabona 4, 70124 Bari, Italy
73: Laboratoire de Physique des 2 infinis, Irene Joliot-Curie,IN2P3/CNRS, Université Paris-Saclay, Université de Paris, 15 rue Georges Clemenceau, 91406 Orsay, Cedex, France
74: INFN Sezione di Pisa, Largo Pontecorvo 3, 56217 Pisa, Italy
75: IRFU/DEDIP, CEA, Université Paris-Saclay, Bat 141, 91191 Gif-sur-Yvette, France

76: INAF - Osservatorio Astronomico di Padova, Vicolo dell'Osservatorio 5, 35122 Padova, Italy
77: INAF - Osservatorio Astronomico di Palermo "G.S. Vaiana", Piazza del Parlamento 1, 90134 Palermo, Italy
78: School of Physics, University of Sydney, Sydney NSW 2006, Australia
79: Sorbonne Université, Université Paris Diderot, Sorbonne Paris Cité, CNRS/IN2P3, Laboratoire de Physique Nucléaire et de Hautes Energies, LPNHE, 4 Place Jussieu, F-75005 Paris, France
80: Instituto de Física de São Carlos, Universidade de São Paulo, Av. Trabalhador São-carlense, 400 - CEP 13566-590, São Carlos, SP, Brazil
81: Departament de Física Quàntica i Astrofísica, Institut de Ciències del Cosmos, Universitat de Barcelona, IEEC-UB, Martí i Franquès, 1, 08028, Barcelona, Spain
82: Department of Physics, Washington University, St. Louis, MO 63130, USA
83: Saha Institute of Nuclear Physics, Bidhannagar, Kolkata-700 064, India
84: INAF - Osservatorio Astronomico di Capodimonte, Via Salita Moiariello 16, 80131 Napoli, Italy
85: Université de Paris, CNRS, Astroparticule et Cosmologie, 10, rue Alice Domon et Léonie Duquet, 75013 Paris Cedex 13, France
86: Astronomy Department of Faculty of Physics, Sofia University, 5 James Bourchier Str., 1164 Sofia, Bulgaria
87: Institut de Recherche en Astrophysique et Planétologie, CNRS-INSU, Université Paul Sabatier, 9 avenue Colonel Roche, BP 44346, 31028 Toulouse Cedex 4, France
88: School of Physics and Astronomy, University of Minnesota, 116 Church Street S.E. Minneapolis, Minnesota 55455-0112, USA
89: IRFU, CEA, Université Paris-Saclay, Bât 141, 91191 Gif-sur-Yvette, France
90: INAF - Istituto di Radioastronomia, Via Gobetti 101, 40129 Bologna, Italy
91: INAF - Istituto di Astrofisica Spaziale e Fisica Cosmica di Palermo, Via U. La Malfa 153, 90146 Palermo, Italy
92: Astronomical Observatory, Department of Physics, University of Warsaw, Aleje Ujazdowskie 4, 00478 Warsaw, Poland
93: Armagh Observatory and Planetarium, College Hill, Armagh BT61 9DG, United Kingdom
94: INFN Sezione di Catania, Via S. Sofia 64, 95123 Catania, Italy
95: INAF - Osservatorio Astronomico di Brera, Via Brera 28, 20121 Milano, Italy
96: Kavli Institute for Particle Astrophysics and Cosmology, Department of Physics and SLAC National Accelerator Laboratory, Stanford University, 2575 Sand Hill Road, Menlo Park, CA 94025, USA
97: Universidade Cruzeiro do Sul, Núcleo de Astrofísica Teórica (NAT/UCS), Rua Galvão Bueno 8687, Bloco B, sala 16, Libertade 01506-000 - São Paulo, Brazil
98: Universidad de Valparaíso, Blanco 951, Valparaiso, Chile
99: INAF - Istituto di Astrofisica e Planetologia Spaziali (IAPS), Via del Fosso del Cavaliere 100, 00133 Roma, Italy
100: Lund Observatory, Lund University, Box 43, SE-22100 Lund, Sweden
101: The Henryk Niewodniczański Institute of Nuclear Physics, Polish Academy of Sciences, ul. Radzikowskiego 152, 31-342 Cracow, Poland
102: Escola de Engenharia de Lorena, Universidade de São Paulo, Área I - Estrada Municipal do Campinho, s/n°, CEP 12602-810, Pte. Nova, Lorena, Brazil
103: INFN Sezione di Trieste and Università degli Studi di Udine, Via delle Scienze 208, 33100 Udine, Italy
104: Palacky University Olomouc, Faculty of Science, RCPTM, 17. listopadu 1192/12, 771 46 Olomouc, Czech Republic
105: Max-Planck-Institut für Physik, Föhringer Ring 6, 80805 München, Germany
106: CENBG, Univ. Bordeaux, CNRS-IN2P3, UMR 5797, 19 Chemin du Solarium, CS 10120, F-33175 Gradignan Cedex, France
107: Dublin City University, Glasnevin, Dublin 9, Ireland
108: Dipartimento di Fisica - Universitá degli Studi di Torino, Via Pietro Giuria 1 - 10125 Torino, Italy
109: Tata Institute of Fundamental Research, Homi Bhabha Road, Colaba, Mumbai 400005, India
110: Universitá degli Studi di Napoli "Federico II" - Dipartimento di Fisica "E. Pancini", Complesso universitario di Monte Sant'Angelo, Via Cintia - 80126 Napoli, Italy
111: Oskar Klein Centre, Department of Physics, University of Stockholm, Albanova, SE-10691, Sweden
112: Yale University, Department of Physics and Astronomy, 260 Whitney Avenue, New Haven, CT 06520-8101, USA
113: CIEMAT, Avda. Complutense 40, 28040 Madrid, Spain
114: University of Oxford, Department of Physics, Denys Wilkinson Building, Keble Road, Oxford OX1 3RH, United Kingdom
115: School of Physics & Astronomy, University of Southampton, University Road, Southampton SO17 1BJ, United Kingdom
116: Department of Physics and Technology, University of Bergen, Museplass 1, 5007 Bergen, Norway
117: Western Sydney University, Locked Bag 1797, Penrith, NSW 2751, Australia
118: School of Physical Sciences, University of Adelaide, Adelaide SA 5005, Australia
119: INFN Sezione di Roma La Sapienza, P.le Aldo Moro, 2 - 00185 Roma, Italy
120: INFN Sezione di Bari, via Orabona 4, 70126 Bari, Italy
121: University of Rijeka, Department of Physics, Radmile Matejcic 2, 51000 Rijeka, Croatia
122: Institute for Theoretical Physics and Astrophysics, Universität Würzburg, Campus Hubland Nord, Emil-Fischer-Str. 31, 97074 Würzburg, Germany
123: Universidade Federal Do Paraná - Setor Palotina, Departamento de Engenharias e Exatas, Rua Pioneiro, 2153, Jardim Dallas, CEP: 85950-000 Palotina, Paraná, Brazil
124: Dept. of Physics and Astronomy, University of Leicester, Leicester, LE1 7RH, United Kingdom
125: Univ. Grenoble Alpes, CNRS, IPAG, 414 rue de la Piscine, Domaine Universitaire, 38041 Grenoble Cedex 9, France

126: National Centre for nuclear research (Narodowe Centrum Badań Jądrowych), Ul. Andrzeja Sołtana7, 05-400 Otwock, Świerk, Poland
127: Enrico Fermi Institute, University of Chicago, 5640 South Ellis Avenue, Chicago, IL 60637, USA
128: Institut für Physik & Astronomie, Universität Potsdam, Karl-Liebknecht-Strasse 24/25, 14476 Potsdam, Germany
129: Department of Physics and Astronomy, Iowa State University, Zaffarano Hall, Ames, IA 50011-3160, USA
130: School of Physics, Aristotle University, Thessaloniki, 54124 Thessaloniki, Greece
131: King's College London, Strand, London, WC2R 2LS, United Kingdom
132: Escola de Artes, Ciências e Humanidades, Universidade de São Paulo, Rua Arlindo Bettio, CEP 03828-000, 1000 São Paulo, Brazil
133: Dept. of Astronomy & Astrophysics, Pennsylvania State University, University Park, PA 16802, USA
134: National Technical University of Athens, Department of Physics, Zografos 9, 15780 Athens, Greece
135: University of Wisconsin, Madison, 500 Lincoln Drive, Madison, WI, 53706, USA
136: Astronomical Observatory of Taras Shevchenko National University of Kyiv, 3 Observatorna Street, Kyiv, 04053, Ukraine
137: Department of Physics, Purdue University, West Lafayette, IN 47907, USA
138: Unitat de Física de les Radiacions, Departament de Física, and CERES-IEEC, Universitat Autònoma de Barcelona, Edifici C3, Campus UAB, 08193 Bellaterra, Spain
139: Institute for Space-Earth Environmental Research, Nagoya University, Chikusa-ku, Nagoya 464-8601, Japan
140: Department of Physical Science, Hiroshima University, Higashi-Hiroshima, Hiroshima 739-8526, Japan
141: Department of Physics, Nagoya University, Chikusa-ku, Nagoya, 464-8602, Japan
142: Friedrich-Alexander-Universität Erlangen-Nürnberg, Erlangen Centre for Astroparticle Physics (ECAP), Erwin-Rommel-Str. 1, 91058 Erlangen, Germany
143: Santa Cruz Institute for Particle Physics and Department of Physics, University of California, Santa Cruz, 1156 High Street, Santa Cruz, CA 95064, USA
144: IRFU / DIS, CEA, Université de Paris-Saclay, Bat 123, 91191 Gif-sur-Yvette, France
145: INFN Sezione di Trieste and Università degli Studi di Trieste, Via Valerio 2 I, 34127 Trieste, Italy
146: School of Physics & Center for Relativistic Astrophysics, Georgia Institute of Technology, 837 State Street, Atlanta, Georgia, 30332-0430, USA
147: Alikhanyan National Science Laboratory, Yerevan Physics Institute, 2 Alikhanyan Brothers St., 0036, Yerevan, Armenia
148: INAF - Telescopio Nazionale Galileo, Roche de los Muchachos Astronomical Observatory, 38787 Garafia, TF, Italy
149: INFN Sezione di Bari and Università degli Studi di Bari, via Orabona 4, 70124 Bari, Italy
150: University of Split - FESB, R. Boskovica 32, 21 000 Split, Croatia
151: Universidad Andres Bello, República 252, Santiago, Chile
152: Academic Computer Centre CYFRONET AGH, ul. Nawojki 11, 30-950 Cracow, Poland
153: University of Liverpool, Oliver Lodge Laboratory, Liverpool L69 7ZE, United Kingdom
154: Department of Physics, Yamagata University, Yamagata, Yamagata 990-8560, Japan
155: Astronomy Department, Adler Planetarium and Astronomy Museum, Chicago, IL 60605, USA
156: Faculty of Management Information, Yamanashi-Gakuin University, Kofu, Yamanashi 400-8575, Japan
157: Department of Physics, Tokai University, 4-1-1, Kita-Kaname, Hiratsuka, Kanagawa 259-1292, Japan
158: Centre for Astrophysics Research, Science & Technology Research Institute, University of Hertfordshire, College Lane, Hertfordshire AL10 9AB, United Kingdom
159: Cherenkov Telescope Array Observatory, Saupfercheckweg 1, 69117 Heidelberg, Germany
160: Tohoku University, Astronomical Institute, Aobaku, Sendai 980-8578, Japan
161: Department of Physics, Rikkyo University, 3-34-1 Nishi-Ikebukuro, Toshima-ku, Tokyo, Japan
162: Department of Physics and Astronomy and the Bartol Research Institute, University of Delaware, Newark, DE 19716, USA
163: Institut für Astro- und Teilchenphysik, Leopold-Franzens-Universität, Technikerstr. 25/8, 6020 Innsbruck, Austria
164: Department of Physics and Astronomy, University of Utah, Salt Lake City, UT 84112-0830, USA
165: IMAPP, Radboud University Nijmegen, P.O. Box 9010, 6500 GL Nijmegen, The Netherlands
166: Josip Juraj Strossmayer University of Osijek, Trg Ljudevita Gaja 6, 31000 Osijek, Croatia
167: Department of Earth and Space Science, Graduate School of Science, Osaka University, Toyonaka 560-0043, Japan
168: Yukawa Institute for Theoretical Physics, Kyoto University, Kyoto 606-8502, Japan
169: Astronomical Observatory, Jagiellonian University, ul. Orla 171, 30-244 Cracow, Poland
170: Landessternwarte, Zentrum für Astronomie der Universität Heidelberg, Königstuhl 12, 69117 Heidelberg, Germany
171: University of Alabama, Tuscaloosa, Department of Physics and Astronomy, Gallalee Hall, Box 870324 Tuscaloosa, AL 35487-0324, USA
172: Department of Physics, University of Bath, Claverton Down, Bath BA2 7AY, United Kingdom
173: University of Iowa, Department of Physics and Astronomy, Van Allen Hall, Iowa City, IA 52242, USA
174: Anton Pannekoek Institute/GRAPPA, University of Amsterdam, Science Park 904 1098 XH Amsterdam, The Netherlands
175: Faculty of Computer Science, Electronics and Telecommunications, AGH University of Science and Technology, Kraków, al. Mickiewicza 30, 30-059 Cracow, Poland
176: Faculty of Science, Ibaraki University, Mito, Ibaraki, 310-8512, Japan
177: Faculty of Science and Engineering, Waseda University, Shinjuku, Tokyo 169-8555, Japan

178: Institute of Astronomy, Faculty of Physics, Astronomy and Informatics, Nicolaus Copernicus University in Toruń, ul. Grudziądzka 5, 87-100 Toruń, Poland
179: Graduate School of Science and Engineering, Saitama University, 255 Simo-Ohkubo, Sakura-ku, Saitama city, Saitama 338-8570, Japan
180: Division of Physics and Astronomy, Graduate School of Science, Kyoto University, Sakyo-ku, Kyoto, 606-8502, Japan
181: Centre for Quantum Technologies, National University Singapore, Block S15, 3 Science Drive 2, Singapore 117543, Singapore
182: Institute of Particle and Nuclear Studies, KEK (High Energy Accelerator Research Organization), 1-1 Oho, Tsukuba, 305-0801, Japan
183: Department of Physics and Astronomy, University of Sheffield, Hounsfield Road, Sheffield S3 7RH, United Kingdom
184: Centro de Ciências Naturais e Humanas, Universidade Federal do ABC, Av. dos Estados, 5001, CEP: 09.210-580, Santo André - SP, Brazil
185: Dipartimento di Fisica e Astronomia, Sezione Astrofisica, Universitá di Catania, Via S. Sofia 78, I-95123 Catania, Italy
186: Department of Physics, Humboldt University Berlin, Newtonstr. 15, 12489 Berlin, Germany
187: Texas Tech University, 2500 Broadway, Lubbock, Texas 79409-1035, USA
188: University of Zielona Góra, ul. Licealna 9, 65-417 Zielona Góra, Poland
189: Institute for Nuclear Research and Nuclear Energy, Bulgarian Academy of Sciences, 72 boul. Tsarigradsko chaussee, 1784 Sofia, Bulgaria
190: University of Białystok, Faculty of Physics, ul. K. Ciołkowskiego 1L, 15-254 Białystok, Poland
191: Faculty of Physics, National and Kapodestrian University of Athens, Panepistimiopolis, 15771 Ilissia, Athens, Greece
192: Universidad de Chile, Av. Libertador Bernardo O'Higgins 1058, Santiago, Chile
193: Hiroshima Astrophysical Science Center, Hiroshima University, Higashi-Hiroshima, Hiroshima 739-8526, Japan
194: Department of Applied Physics, University of Miyazaki, 1-1 Gakuen Kibana-dai Nishi, Miyazaki, 889-2192, Japan
195: School of Allied Health Sciences, Kitasato University, Sagamihara, Kanagawa 228-8555, Japan
196: Departamento de Astronomía, Universidad de Concepción, Barrio Universitario S/N, Concepción, Chile
197: Charles University, Institute of Particle & Nuclear Physics, V Holešovičkách 2, 180 00 Prague 8, Czech Republic
198: Astronomical Observatory of Ivan Franko National University of Lviv, 8 Kyryla i Mephodia Street, Lviv, 79005, Ukraine
199: Kobayashi-Maskawa Institute (KMI) for the Origin of Particles and the Universe, Nagoya University, Chikusa-ku, Nagoya 464-8602, Japan
200: Graduate School of Technology, Industrial and Social Sciences, Tokushima University, Tokushima 770-8506, Japan
201: Space Research Centre, Polish Academy of Sciences, ul. Bartycka 18A, 00-716 Warsaw, Poland
202: Instituto de Física - Universidade de São Paulo, Rua do Matão Travessa R Nr.187 CEP 05508-090 Cidade Universitária, São Paulo, Brazil
203: International Institute of Physics at the Federal University of Rio Grande do Norte, Campus Universitário, Lagoa Nova CEP 59078-970 Rio Grande do Norte, Brazil
204: University College Dublin, Belfield, Dublin 4, Ireland
205: Centre for Astro-Particle Physics (CAPP) and Department of Physics, University of Johannesburg, PO Box 524, Auckland Park 2006, South Africa
206: Departamento de Física, Facultad de Ciencias Básicas, Universidad Metropolitana de Ciencias de la Educación, Santiago, Chile
207: Núcleo de Formação de Professores - Universidade Federal de São Carlos, Rodovia Washington Luís, km 235 CEP 13565-905 - SP-310 São Carlos - São Paulo, Brazil
208: Physik-Institut, Universität Zürich, Winterthurerstrasse 190, 8057 Zürich, Switzerland
209: Department of Physical Sciences, Aoyama Gakuin University, Fuchinobe, Sagamihara, Kanagawa, 252-5258, Japan
210: University of the Free State, Nelson Mandela Avenue, Bloemfontein, 9300, South Africa
211: Faculty of Electronics and Information, Warsaw University of Technology, ul. Nowowiejska 15/19, 00-665 Warsaw, Poland
212: Rudjer Boskovic Institute, Bijenicka 54, 10 000 Zagreb, Croatia
213: Department of Physics, Konan University, Kobe, Hyogo, 658-8501, Japan
214: Kumamoto University, 2-39-1 Kurokami, Kumamoto, 860-8555, Japan
215: University School for Advanced Studies IUSS Pavia, Palazzo del Broletto, Piazza della Vittoria 15, 27100 Pavia, Italy
216: Aalto University, Otakaari 1, 00076 Aalto, Finland
217: Agenzia Spaziale Italiana (ASI), 00133 Roma, Italy
218: Observatoire de la Cote d'Azur, Boulevard de l'Observatoire CS34229, 06304 Nice Cedex 4, Franc